\documentclass[twocolumn,trackchanges]{aastex701}

\usepackage{soul}

\usepackage{multirow}
\usepackage{amsmath}
\usepackage{enumerate}
\begin{document}

\title{The FAST Galactic Plane Pulsar Snapshot Survey. IX. Timing three binary pulsars with wide orbits and low orbital eccentricities}

\author[orcid=0009-0009-6590-1540]{Z.~L. Yang}
\affiliation{National Astronomical Observatories, Chinese Academy of Sciences, Jia-20 Datun Road, ChaoYang District, Beijing 100012, China}
\affiliation{School of Astronomy and Space Science, University of Chinese Academy of Sciences, Beijing 100049, China}
\email{zlyang@nao.cas.cn}  

\author[orcid=0000-0002-9274-3092]{J.~L. Han} 
\affiliation{National Astronomical Observatories, Chinese Academy of Sciences, Jia-20 Datun Road, ChaoYang District, Beijing 100012, China}
\affiliation{School of Astronomy and Space Science, University of Chinese Academy of Sciences, Beijing 100049, China}
\affiliation{State Key Laboratory of Radio Astronomy and Technology, Beijing 100101, China }
\email[show]{hjl@nao.cas.cn}  

\author[orcid=0009-0003-2212-4792]{W.~Q. Su}
\affiliation{National Astronomical Observatories, Chinese Academy of Sciences, Jia-20 Datun Road, ChaoYang District, Beijing 100012, China}
\affiliation{School of Astronomy and Space Science, University of Chinese Academy of Sciences, Beijing 100049, China}
\email{suwq@nao.cas.cn}

\author[orcid=0009-0004-3433-2027]{C. Wang}
\affiliation{National Astronomical Observatories, Chinese Academy of Sciences, Jia-20 Datun Road, ChaoYang District, Beijing 100012, China}
\affiliation{School of Astronomy and Space Science, University of Chinese Academy of Sciences, Beijing 100049, China}
\affiliation{State Key Laboratory of Radio Astronomy and Technology, Beijing 100101, China }
\email{wangcheni@nao.cas.cn}

\author[]{J.~P. Yuan}
\affiliation{Xinjiang Astronomical Observatory, Chinese Academy of Sciences, Urumqi 830011, China}
\affiliation{State Key Laboratory of Radio Astronomy and Technology, Beijing 100101, China }
\email{yuanjp@xao.ac.cn}

\author[orcid=0000-0002-4704-5340]{T. Wang}
\affiliation{National Astronomical Observatories, Chinese Academy of Sciences, Jia-20 Datun Road, ChaoYang District, Beijing 100012, China}
\email{twang@nao.cas.cn}

\author[orcid=0009-0008-1612-9948]{Yi Yan}
\affiliation{National Astronomical Observatories, Chinese Academy of Sciences, Jia-20 Datun Road, ChaoYang District, Beijing 100012, China}
\affiliation{School of Astronomy and Space Science, University of Chinese Academy of Sciences, Beijing 100049, China}
\email{yanyi@nao.cas.cn}
 
\author[orcid=0000-0003-1778-5580]{J. Xu}
\affiliation{National Astronomical Observatories, Chinese Academy of Sciences, Jia-20 Datun Road, ChaoYang District, Beijing 100012, China}
\affiliation{State Key Laboratory of Radio Astronomy and Technology, Beijing 100101, China }
\email{xujun@nao.cas.cn}

\author[orcid=0000-0002-1056-5895]{W.~C. Jing}
\affiliation{National Astronomical Observatories, Chinese Academy of Sciences, Jia-20 Datun Road, ChaoYang District, Beijing 100012, China}
\affiliation{School of Astronomy and Space Science, University of Chinese Academy of Sciences, Beijing 100049, China}
\email{wcjing@nao.cas.cn}

\author[orcid=0000-0002-6437-0487]{P.~F. Wang}
\affiliation{National Astronomical Observatories, Chinese Academy of Sciences, Jia-20 Datun Road, ChaoYang District, Beijing 100012, China}
\affiliation{School of Astronomy and Space Science, University of Chinese Academy of Sciences, Beijing 100049, China}
\affiliation{State Key Laboratory of Radio Astronomy and Technology, Beijing 100101, China }
\email{pfwang@nao.cas.cn}

\author[orcid=0000-0002-5915-5539]{N.~N. Cai}
\affiliation{National Astronomical Observatories, Chinese Academy of Sciences, Jia-20 Datun Road, ChaoYang District, Beijing 100012, China}
\email{cnn@nao.cas.cn}

\author[orcid=0000-0002-6423-6106]{D.~J. Zhou}
\affiliation{National Astronomical Observatories, Chinese Academy of Sciences, Jia-20 Datun Road, ChaoYang District, Beijing 100012, China}
\email{djzhou@nao.cas.cn}

\author[]{X.~J. Chen}
\affiliation{Xinjiang Astronomical Observatory, Chinese Academy of Sciences, Urumqi 830011, China}
\email{chenxingjiang@xao.ac.cn}

\author[]{D. Zhao}
\affiliation{Xinjiang Astronomical Observatory, Chinese Academy of Sciences, Urumqi 830011, China}
\email{zhaode@xao.ac.cn}

\begin{abstract}
Current pulsar timing models face challenges when applied to binary pulsars with wide orbits and low orbital eccentricities. The conventional \texttt{DD} model accurately characterizes the orbits of such systems, but it suffers from strong correlations between the time of periastron passage ($T_0$) and the longitude of periastron ($\omega$). The \texttt{ELL1} model avoids these parameter correlations, yet fails due to the limitations of its first-order low-eccentricity approximation. Recent enhancements to the \texttt{ELL1} model (dubbed \texttt{ELL1+} model) have incorporated higher-order terms but retain the low-eccentricity approximation. In this study, we propose a further improved model, \texttt{ELL1R}, which eliminates reliance on the low-eccentricity approximation through rigorous calculation of the R\"{o}mer delay.  This modification can avoid strong parameter correlations in the \texttt{DD} model, and it can be used in systems with mild eccentricity $0.01\lesssim e\lesssim0.1$ where the \texttt{ELL1+} model can not. Using the \texttt{ELL1R} model, we present the first phase-coherent timing solutions for three binary pulsars: PSR~J1851--0108 (orbital period: 228 days), PSR~J1910+0423 (886 days), and PSR~J1923+2022 (777 days). Validation against the \texttt{DD} and \texttt{ELL1+} models confirms that \texttt{ELL1R} yields consistent timing results while integrating the advantages of the two models. Our analysis further indicates that all three pulsars are mildly recycled. The companions of PSRs J1910+0423 and J1923+2022 are likely white dwarfs, whereas the nature of PSR J1851--0108’s companion remains unknown.
\end{abstract}



\section{Introduction} 

Binary pulsars have various types of companions with very different orbital parameters. Pulsar-white dwarf systems have small orbital eccentricities, and their orbital periods can be as long as about 1000 days \citep{Manchester+2005AJ....129.1993M}. These binary systems allow us to constrain  violations of the strong equivalence principle and  momentum conservation as well as the Lorentz invariance \citep{Damour+1991PhRvL..66.2549D, Wex+1997A&A...317..976W, Wex+2000ASPC..202..113W, Will+1993tegp.book.....W}.
Parameters of pulsar timing solutions contain important information on locations, rotational slowdown, binary orbits, and relativistic effects. 
By comparing the predictions of the timing solutions with the measured times of pulse arrival (TOAs), the values and uncertainties of these parameters can be obtained.

Many timing models have been developed for binary pulsar timing, and selecting an appropriate model is essential for achieving a reliable phase-connected timing solution. The \texttt{BT} model \citep{Blandford+1976ApJ...205..580B} was proposed for the timing of the first known binary pulsar, PSR B1913+16 \citep{Hulse+1975ApJ...195L..51H}.  Later, \citet{Damour+1985AIHPA..43..107D, Damour+1986AIHPA..44..263D} developed the \texttt{DD} model which provides a better description of the relativistic effects, and nowadays the \texttt{DD} model is widely employed for binary pulsars with significant orbital eccentricities, such as double neutron star systems. However, for binary pulsars with low eccentricity ($e \ll 1$), the parameters $\omega$ (the longitude of periastron) and $T_0$ (the time of periastron passage) exhibit strong covariance.  
When fitting the orbital eccentricity $e$ in such systems, timing software may terminate due to $e < 0$, complicating the search for a phase-connected solution.

\begin{table*}[ht]
    \footnotesize
    \caption{FAST observation sessions for three pulsars.}
    \centering
    \footnotesize
    \setlength\tabcolsep{3pt}
    \begin{tabular}{lccc}
    \hline
     PSR   & J1851--0108 & J1910+0423 & J1923+2022 \\ \hline \hline 
     Observation MJD range & 59480 to 61005   & 59768 to 61014 & 59180 to 60999 \\
    Total sessions & 28 & 21 & 33 \\
      \hline
    FAST GPPS (PI: J.~L. Han)   & 4 & 9 & 6 \\
    PT2022\_0047 (PI: W.~Q. Su) &17 & 0 &14 \\
    PT2023\_0017 (PI: Chen Wang) & 0 & 0 & 3 \\
    PT2023\_0102 (PI: W.~Q.~Su) & 0 & 1 & 0 \\
    PT2023\_0162 (PI: W.~Q.~Su)& 2 & 3 & 1 \\
    PT2024\_0089 (PI: Yi Yan) & 0 & 2 & 0 \\
    PT2024\_0173 (PI: Jun Xu) & 0 & 0 & 1 \\
    PT2024\_0195 (PI: Chen Wang) & 0 & 0 & 1 \\
    PT2024\_0199 (PI: W.~Q.~Su) & 3 & 3 & 1 \\   
    PT2024\_0235 (PI: W.~C. Jing)& 1 & 0 & 0 \\
    ZD2025\_6 (PI: J.~P.~Yuan) & 0 & 2 & 6 \\
    PT2025\_0211 (PI: T.~Wang) & 0 & 1 & 0 \\
    PT2025\_0221 (PI: W.~Q.~Su) & 1 & 0 & 0 \\
    \hline
    \end{tabular}
    \label{FASTobs}
\end{table*}

For low-eccentricity binary pulsars, such as many pulsar–white dwarf systems, \citet{Lange+2001MNRAS.326..274L} introduced the \texttt{ELL1} timing model, in which $\omega$, $T_0$, and $e$ are replaced by the time of ascending node passage $T_{\rm asc} \equiv T_0 - \omega/2\pi P_{\rm orb}$ and two Laplace parameters $\epsilon_1 \equiv e \sin\omega$ and $\epsilon_2 \equiv e \cos\omega$. 
%
The R\"{o}mer delay $\Delta_{\rm RB}$, which describes the variation in light travel time caused by orbital motion, is computed only to first order in the orbital eccentricity $e$ \citep{Lange+2001MNRAS.326..274L}. This approximation holds when $xe^2/c \lesssim \sigma_{\rm TOA}/N^{1/2}_{\rm TOA}$, where $x$ is the projected semi-major axis, $c$ is the speed of light, $\sigma_{\rm TOA}$ is the  uncertainty of individual TOA measurements, and $N_{\rm TOA}$ is the number of TOAs. However, in many wide-orbit pulsar–white dwarf binary systems, $xe^2$ can reach values $\gtrsim$1 ms, and 
the accuracy of the first-order approximation is not enough.

Subsequent improvements were introduced by \citet{Zhu+2019MNRAS.482.3249Z}, who derived the second-order eccentricity terms, and later by \citet{Fiore+2023ApJ...956...40F}, who extended the calculation to third order. This enhanced version, referred to hereafter as the \texttt{ELL1+} model, has been implemented in the \textsc{Tempo}\footnote{http://tempo.sourceforge.net/} software \citep[including terms up to $xe^2$;][]{Nice+2015ascl.soft09002N} and the \textsc{Pint} pulsar timing package \citep[up to $xe^3$;][]{Luo+2021ApJ...911...45L}, though it remains unavailable in the widely used \textsc{Tempo2} software \citep{Hobbs+2006MNRAS.369..655H}.

In this paper, we present a refined timing model \citep[hereafter referred to as the \texttt{ELL1R} model\footnote{available in https://github.com/Zony-naoc/ELL1Rmodel};][]{ell1r}  written in \textsc{Tempo2} format \citep{Hobbs+2006MNRAS.369..655H} that computes the R\"{o}mer delay without relying on low-eccentricity approximations. We then applied it to three binary systems, PSRs J1851--0108, J1910+0423, and J1923+2022, 
which were discovered by the Five-hundred-meter Aperture Spherical radio Telescope \citep[FAST, ][]{Nan2006} during the Galactic Plane Pulsar Snapshot (GPPS) survey \citep{Han+2021RAA....21..107H, Han+2025RAA....25a4001H}. The initial orbital parameters of PSRs J1910+0423 and J1923+2022 have been published in \citet{Wang+2025RAA....25a4003W}, and they are in orbits with long orbital periods of a few hundred days. PSRs~J1910+0423 and J1923+2022 have very small orbital eccentricities $e<0.01$, while PSRs J1851--0108 has a mild orbital eccentricity $0.01\lesssim e\lesssim0.1$. We obtained their new timing solutions accurately.
In Section 2, FAST observations, the timing procedure, and our new timing model are introduced. In Section 3, we present the timing results of PSRs J1851--0108, J1910+0423, and J1923+2022 by using different timing models and discuss the pulsar properties and their possible origins. The summary of our work is presented in Section 4.

\section{Observations and Data Reduction}

\subsection{FAST Observations}

The three pulsars, 
PSRs J1851--0108, J1910+0423 and J1923+2022 were discovered in the FAST GPPS survey \citep{Han+2021RAA....21..107H, Han+2025RAA....25a4001H}. 
PSR~J1851--0108 was discovered in a 21-min snapshot observation on 2021 September 23 and later confirmed by a 15-min tracking observation on 2021 October 5; PSR~J1910+0423 was discovered in a 21-min snapshot observation on 2022 July 9 and later confirmed by a 5-min tracking observation on 2022 July 22; PSR~J1923+2022 was discovered in a 15-min tracking observation on 2020 November 27 and later confirmed by another 15-min tracking observation on 2021 January 27. Their binary characteristics have been identified using changes of their barycentric spin periods \citep{Wang+2025RAA....25a4003W}.

In this work, we 
used all available FAST observations from the FAST GPPS survey 
and other FAST projects (see Table~\ref{FASTobs}). 
The $L$-band 19-beam receiver has been used for all FAST observations, which covers the frequency range of 1.0 to 1.5 GHz. Data were recorded in the search mode with 4096 or 2048 frequency channels, with a sampling time of 49.152 $\mu$s. In most observations, the data from 4 polarization channels from the linear polarization feed, $XX$, $YY$, Re[$X^{*}Y$] and Im[$X^{*}Y$], were recorded. The polarization data were calibrated by using the periodic calibration noise signals injected for 120 s (for 10-min or longer sessions) or 40 s (for other sessions) at the beginning or the end of each observation session. 

\begin{figure*}[ht]
    \centering
    \includegraphics[width=0.75\linewidth]{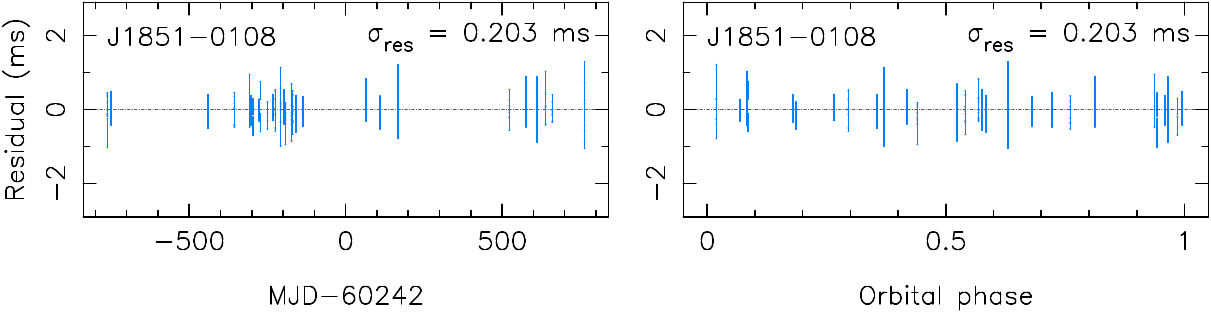}\\[4mm]
    \includegraphics[width=0.75\linewidth]{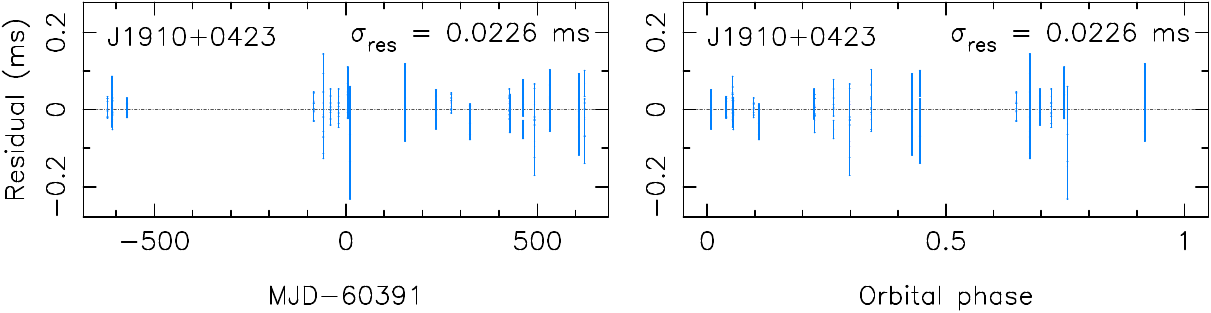}\\[4mm]
    \includegraphics[width=0.75\linewidth]{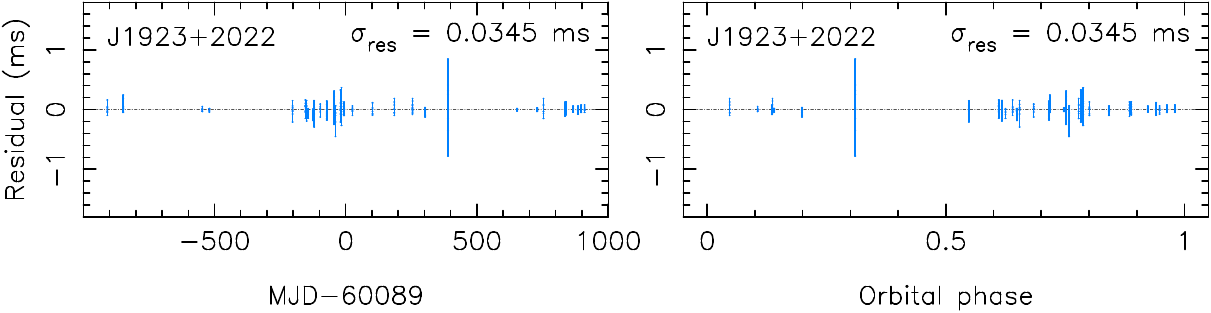}
    \caption{Timing residuals of PSRs J1851--0108, J1910+0423 and J1923+2022 as as a function of epoch or orbital phase. The error bars show the $1\sigma$ uncertainties of the measured TOAs taken with the $L$-band 19-beam receiver of FAST. No systematic trends are visible. Their weighted root-mean-square of timing residuals $\sigma_{\rm res}$ are shown in each subpanel.}
    \label{timing_res}
\end{figure*}

\subsection{Pulsar timing procedure and \texttt{ELL1R} timing model}

The orbital motion of a binary pulsar gives rise to variations in its line-of-sight velocity, which in turn cause changes in the observed barycentric spin period. These variations are described by a set of Keplerian orbital parameters: the orbital period $P_{\rm orb}$, the time of periastron passage $T_0$, the longitude of periastron $\omega$, the projected semi-major axis $x$, and the orbital eccentricity $e$. For each observation session, we derived the barycentric spin period $P_{\rm bary}$ and the barycentric epoch $T_{\rm bary}$ using the \texttt{prepfold} command in \textsc{Presto}, a software package designed for pulsar data analysis \citep{Ransom+2001PhDT.......123R}. Adopting the approach of \citet{Bhattacharyya+2008MNRAS.387..273B}, we then performed a two- or four-dimensional search over orbital parameters to fit the barycentric periods measured at different epochs.

Using these initial orbital parameters, we folded the observational data into pulse profiles using \textsc{Dspsr} \citep{Straten+2011PASA...28....1V}. Radio frequency interference (RFI) was automatically removed using the method described in \citet{Chen+2023RAA....23j4004C}, and polarization profiles were derived following the procedure in \citet{Wang+2023RAA....23j4002W}. TOAs were extracted from the resulting profiles using the \textsc{Psrchive} software suite through the following steps: (1) manual verification of RFI excision using \texttt{psrzap}; (2) compressing each session’s data into one polarization channel (total intensity), two sub-integrations, and four sub-bands using the \texttt{pam} command \citep{Hotan+2004PASA...21..302H}; and (3) extracting TOAs via the \texttt{pat} command by cross-correlating the profiles with a one-dimensional frequency-averaged template generated by \texttt{paas} \citep{Hotan+2004PASA...21..302H}.

\begin{table*}[ht]
    \caption{Timing model parameters and other important parameters for PSRs J1851--0108, J1910+0423 and J1923+2022 in TDB time scale at reference epoch MJD 60500. Distances are estimated using the NE2001 \citep{Cordes+2002astro.ph..7156C} and YMW16 \citep{Yao+2017ApJ...835...29Y} electron-density models. The Solar System ephemeris used in this work is DE440 \citep{Park+2021AJ....161..105P}.}
    \centering
    \footnotesize
    \renewcommand\arraystretch{0.9}{
    \begin{tabular}{lccc}
    \hline
    Parameters  &  J1851--0108 & J1910+0423 & J1923+2022\\ \hline
     $N_{\rm TOA}$& 98 & 82 & 131\\
     $\chi^2_{\rm min}/N_{\rm free}$ & 0.8249 & 1.1775 & 1.2500 \\[1mm]
    \hline
    Right ascension, $\alpha$ (J2000 equinox) & $18^{\rm h}51^{\rm m}11^{\rm s}.390(2)$ & $19^{\rm h}10^{\rm m}10^{\rm s}.2275(15)$ & $19^{\rm h}23^{\rm m}47^{\rm s}.4763(5)$ \\
    Declination, $\delta$ (J2000 equinox)   & -$01^{\circ}08'10''.04(8)$ & +$04^\circ23'01''.19(3)$ & +$20^\circ22'50''.676(11)$ \\
    Spin period, $P$ (s) & 0.08707461410585(12) & 0.0932414759973(5) &0.037992888786748(15)\\
    Spin period derivative, $\dot{P}$ (s s$^{-1}$) & 2.75(10)$\times10^{-19}$ & 9.2(5)$\times10^{-19}$ &1.812(2)$\times10^{-19}$\\
    Spin frequency, $\nu$ (Hz) & 11.484403465565(16) & 10.72484094985(5) &26.320715058361(10) \\
    Spin frequency derivative, $\dot{\nu}$ (s$^{-2}$) & -3.63(13)$\times10^{-17}$& -1.06(5)$\times10^{-16}$ & -1.2552(14)$\times10^{-16}$\\
    Characteristic age,  $\tau_{\rm c}$ (Gyr)      &  5.0 & 1.6 & 3.3 \\
    Surface magnetic field strength, $B_{\rm s}$ (G)  &   5.0$\times$10$^9$& 9.4$\times$10$^9$ & 2.7$\times$10$^9$\\
    Dispersion measure (pc cm$^{-3}$) & 670.96(5) & 339.840(5) & 175.310(6) \\
    Dispersion measure derivative (pc cm$^{-3}$ yr$^{-1}$) & - & 0.016(4) & 0.007(4) \\
    Distance from NE2001 \& YMW16 (kpc)  & 8.8\&6.1 & 8.4\&13.1 & 6.2\&4.6 \\   
    Rotation measure (rad\,m$^{-2}$) & 319(2) & 634(2) & 223(4) \\
    \hline
    Binary model & \multicolumn{3}{c}{\texttt{DD}} \\
    \hline
    Orbital period, $P_{\rm orb}$ (day) & 228.021994(9) & 885.8686(4) & 777.470128(8)\\
    Projected semi-major axis, $x$ (lt-s) & 207.92211(5) & 203.3159(2) & 244.613429(15) \\
    Time of periastron passage, $T_0$ (MJD) & 60267.67630(12) & 61243.94(3) & 60493.53(3)\\
    Longitude of periastron, $\omega$ (deg) & 142.23572(19) & 253.885(11) & 118.480(13) \\
    Orbital eccentricity, $e$ & 0.1331659(4) & 0.0026133(8) & 0.00067731(18) \\
    Weight rms of timing residuals, $\sigma_{\rm res}$ (\textmu s) & 203.142 & 22.582 & 34.540 \\
    \hline
    Binary model & \multicolumn{3}{c}{\texttt{ELL1+}} \\
    \hline
    Orbital period, $P_{\rm orb}$ (day) & -& 885.8686(4) & 777.470128(8)\\
    Projected semi-major axis, $x$ (lt-s) & - & 203.3159(2) & 244.613429(15) \\
    Time of ascending node passage, $T_{\rm asc}$ (MJD) & - & 60619.19547(12)  & 60237.658307(9) \\  
    First Laplace parameter, $\epsilon_1\equiv e\sin\omega$  &  - & -0.0025106(8) & 0.00059534(15)\\
    Second Laplace parameter,  $\epsilon_2\equiv e\cos\omega$ & - & -0.0007254(5) & -0.00032298(19)  \\
    Weighted rms of timing residuals, $\sigma_{\rm res}$ (\textmu s) & - & 22.582 & 34.541 \\
    \hline
    Binary model & \multicolumn{3}{c}{\texttt{ELL1R}} \\
    \hline
    Orbital period, $P_{\rm orb}$ (day) & 228.021994(9) & 885.8686(4) & 777.470128(8)\\
    Projected semi-major axis, $x$ (lt-s) & 207.92211(5) & 203.3159(2) & 244.613429(15) \\
    Time of ascending node passage, $T_{\rm asc}$ (MJD) & 60177.584981(9) & 60619.19546(12) & 60237.658309(9) \\  
    First Laplace parameter, $\epsilon_1\equiv e\sin\omega$  &  0.0815527(3) & -0.0025106(8) &0.00059534(15)  \\
    Second Laplace parameter,  $\epsilon_2\equiv e\cos\omega$ & -0.1052726(5) & -0.0007254(5) &-0.00032298(19)  \\
    Weight rms of timing residuals, $\sigma_{\rm res}$ (\textmu s) & 203.142 & 22.582 & 34.540 \\
    \hline
    Binary model & \multicolumn{3}{c}{Derived binary parameters from \texttt{ELL1R} model}\\
    \hline
    Time of periastron passage, $T_0$ (MJD) & 60267.67630(11) & 61243.94(3) & 60493.53(3)\\
    Longitude of periastron, $\omega$ (deg) & 142.23572(17) & 253.885(12) & 118.480(15) \\
    Orbital eccentricity, $e$ & 0.1331659(4) & 0.0026133(8) & 0.00067731(16) \\
    Mass function ($M_\odot$) & 0.18562277(14) & 0.01149896(2) & 0.025998973(4)\\
    Minimum companion mass ($M_\odot$) & 1.03 & 0.32 & 0.45\\
    \hline
    \end{tabular}
    }
    \label{basicparameter}
\end{table*}

The TOAs were analyzed using the high-precision pulsar timing package \textsc{Tempo2} \citep{Hobbs+2006MNRAS.369..655H}. The \texttt{ELL1} timing model is commonly used for binary systems with low orbital eccentricity. In this model, the R\"{o}mer delay $\Delta_{\rm RB}$ is given by \citep{Lange+2001MNRAS.326..274L}:
\begin{equation}
    \Delta_{\rm RB} \approx x\left[ \sin\Phi + \frac{\epsilon_1}{2} \sin(2\Phi) - \frac{\epsilon_2}{2} \cos(2\Phi) \right],
\end{equation}
where $\Phi$ is the orbital phase. It is worth noting that, in addition to terms of order $xe^2$, the \texttt{ELL1} model also omits the term $xe \sin\omega \equiv x\epsilon_1$ in the R\"{o}mer delay expression \citep{Susobhanan+2018MNRAS.480.5260S}. If $x\epsilon_1$ varies with time, this omission can introduce a small systematic bias in the measured spin period.
For wide-orbit systems  PSRs J1851--0108, J1910+0423, and J1923+2022 whose $xe^2/c$ are 3.7, 1.4$\times10^{-3}$ and 1.1$\times10^{-4}$ s, respectively, the original \texttt{ELL1} model implemented in \textsc{Tempo2} is not well-suited for their timing since the low-eccentricity approximation $xe^2 \lesssim \sigma_{\rm TOA}/N^{1/2}_{\rm TOA}$ is no longer true. The \texttt{ELL1+} model can be used for timing PSRs J1910+0423 and J1923+2022, but it remains unsuitable for PSR J1851--0108 due to its large $xe^4/c=0.065$ s. Note that the ELL1+ model is not currently included in the standard \textsc{Tempo2} distribution  \citep{Hobbs+2006MNRAS.369..655H}.

Although the \texttt{BT} and \texttt{DD} models can accurately calculate the R\"{o}mer delay, they are not suitable for systems with low orbital eccentricities. Traditionally, the last digit of a measured value is uncertain and should align with the place of the uncertainty. In the \texttt{BT} or \texttt{DD} model, the high covariance between $\omega$ and $T_0$ significantly increases their uncertainties. Therefore, if the \texttt{BT} or \texttt{DD} model was applied for these systems, the reported values of $\omega$ and $T_0$ cannot accurately reproduce the timing results. Besides that, the very low orbital eccentricities of PSRs J1910+0423 and J1923+2022 make finding a phase-connected timing solution much more difficult, as timing software may terminate due to $e < 0$.

To address this issue, we developed a modified timing model implemented in \textsc{Tempo2} \citep{Hobbs+2006MNRAS.369..655H}, \texttt{ELL1R}, in which the parameters $T_{\rm asc}$, $\epsilon_1$, and $\epsilon_2$ are converted back into $T_0$, $e$, and $\omega$ as follows:
\begin{equation}
\begin{split}
T_0 &= T_{\rm asc} + \frac{\arctan(\epsilon_1 / \epsilon_2)}{2\pi}P_{\rm orb}, \\
e &= \sqrt{\epsilon_1^2 + \epsilon_2^2}, \\
\omega &= \arctan(\epsilon_1 / \epsilon_2).
\end{split}
\label{T0ECCOM}
\end{equation}
The R\"{o}mer delay is then computed according to the \texttt{BT} and \texttt{DD} models \citep{Blandford+1976ApJ...205..580B,Damour+1985AIHPA..43..107D, Damour+1986AIHPA..44..263D}:
\begin{equation}
\Delta_{\rm RB} = x \sin\omega (\cos u - e) + x \sqrt{1 - e^2} \cos\omega \sin u,
\label{Romer_delay}
\end{equation}
where $u$ is the eccentric anomaly. This original-sense recovered model, dubbed the \texttt{ELL1R} model, can be applied to most systems with $e \ll 1$ --- including those in wide orbits with mild eccentricities of $0.01\lesssim e\lesssim0.1$ for which the \texttt{ELL1+} model is not suitable --- and avoids the strong correlation between $T_0$ and $\omega$ in the \texttt{BT} or \texttt{DD} model.

In the \texttt{ELL1} model \citep{Lange+2001MNRAS.326..274L}, the Shapiro delay $\Delta_{\rm SB}$  is approximated as:
\begin{equation}
    \Delta_{\rm SB} \approx -\frac{2G m_{\rm c}}{c^3}\ln(1-\sin i\sin\Phi),
\end{equation}
where $G$ is the gravitational constant, $m_{\rm c}$ is the companion mass, and $i$ is the orbital inclination. Following the \texttt{DD} model \citep{Damour+1986AIHPA..44..263D}, we strictly compute the Shapiro delay in the \texttt{ELL1R} model:
\begin{equation}
\begin{split}
    \Delta_{\rm SB} = & -\frac{2G m_{\rm c}}{c^3}
\ln\Bigr\{1  - \sin i \Big[ \sin \omega (\cos u - e) \\
& \left.\left. + \sqrt{1 - e^2} \cos \omega \sin u \right] - e \cos u\right\}.
    \label{Delta_S}
\end{split}
\end{equation}

Currently, aside from the R\"{o}mer delay and the Shapiro delay, following the origin \texttt{ELL1} model \citep{Lange+2001MNRAS.326..274L} the \texttt{ELL1R} model only accounts for secular variations in the orbital parameters. Consequently, for compact binary pulsars with mild eccentricities such as the double pulsar system PSR~J0737$-$3039A/B \citep{Lyne+2004Sci...303.1153L}, the \texttt{DD} model is still required to accurately describe the relativistic effects \citep{Damour+1985AIHPA..43..107D, Damour+1986AIHPA..44..263D}.

In estimating parameter uncertainties, \textsc{Tempo2} \citep{Hobbs+2006MNRAS.369..655H} computes the covariance matrix $C$ from the first‑order partial derivatives. Using Eq.~\ref{T0ECCOM}, we transform the partial derivatives with respect to $T_0$, $\omega$, and $e$ into those with respect to $T_{\rm asc}$, $\epsilon_1$, and $\epsilon_2$. The correlation coefficient $r$ between fitted parameter 1 and fitted parameter 2 is then given by $r=C_{12}/\sqrt{C_{11}C_{22}}$. Even for a mildly eccentric system such as PSR~J1851--0108, the \texttt{DD} model \citep{Damour+1985AIHPA..43..107D, Damour+1986AIHPA..44..263D} yields a correlation coefficient between $\omega$ and $T_0$ as high as 0.998. In contrast, the \texttt{ELL1R} model avoids such a strong parameter correlation. For PSR~J1851--0108, the highest correlation coefficient among the three orbital parameters adopted in the \texttt{ELL1R} model --- $T_{\rm asc}$, $\epsilon_1$, and $\epsilon_2$ --- is 0.366, observed between $T_{\rm asc}$ and $\epsilon_2$. The highest correlation coefficient among all fitting parameters is -0.887, observed between $P_{\rm orb}$ and the spin frequency derivative $\dot{\nu}$. The correlation arises from limited observational coverage, an issue that is also present in the \texttt{DD} model.

\begin{figure*}[tbh]
    \centering
    \includegraphics[width=0.32\textwidth]{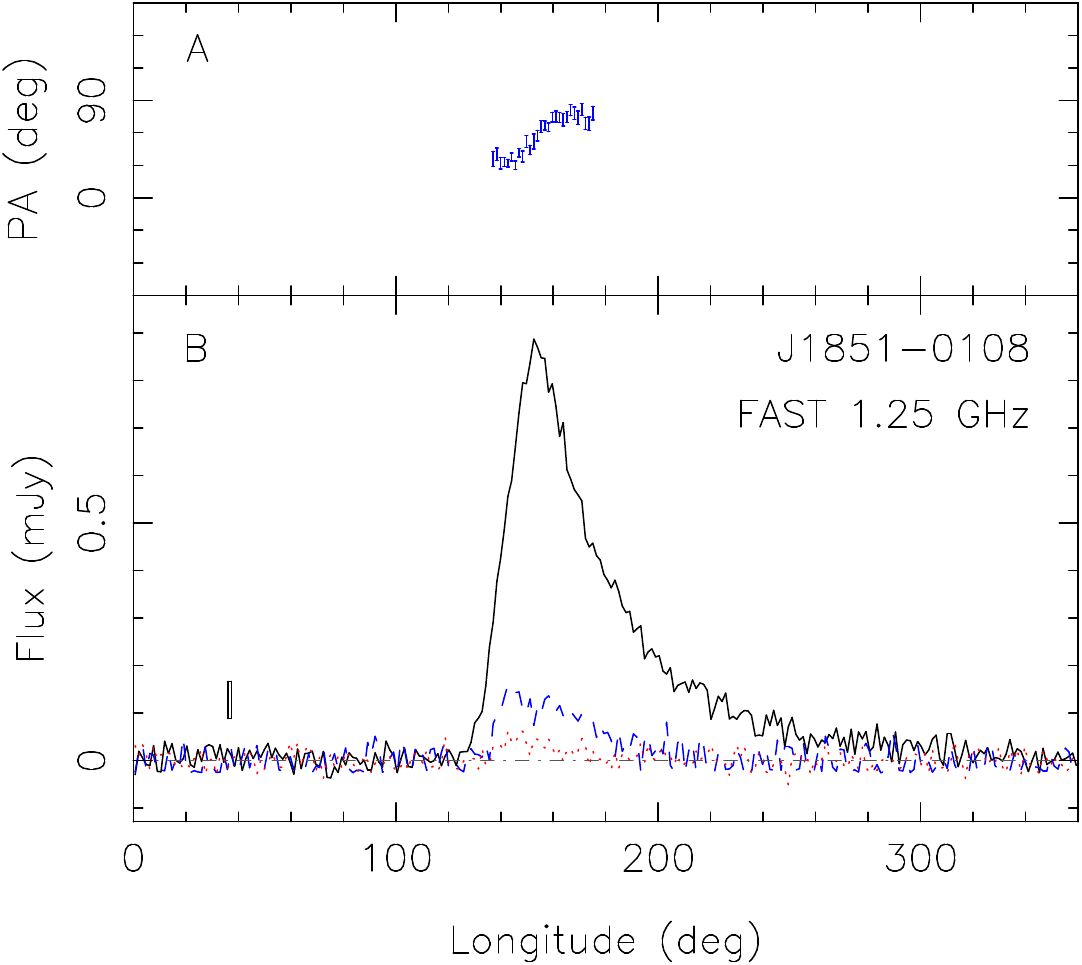}
    \includegraphics[width=0.32\textwidth]{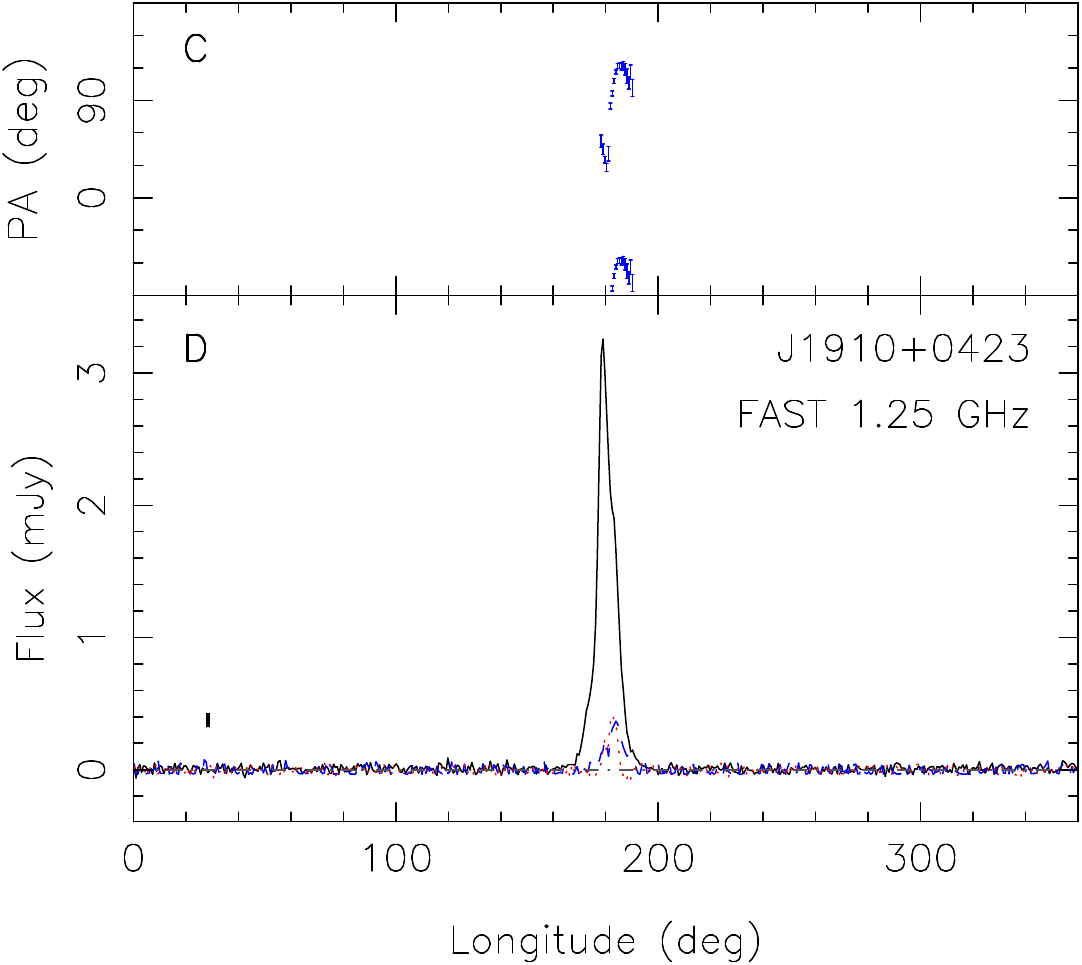}
    \includegraphics[width=0.32\textwidth]{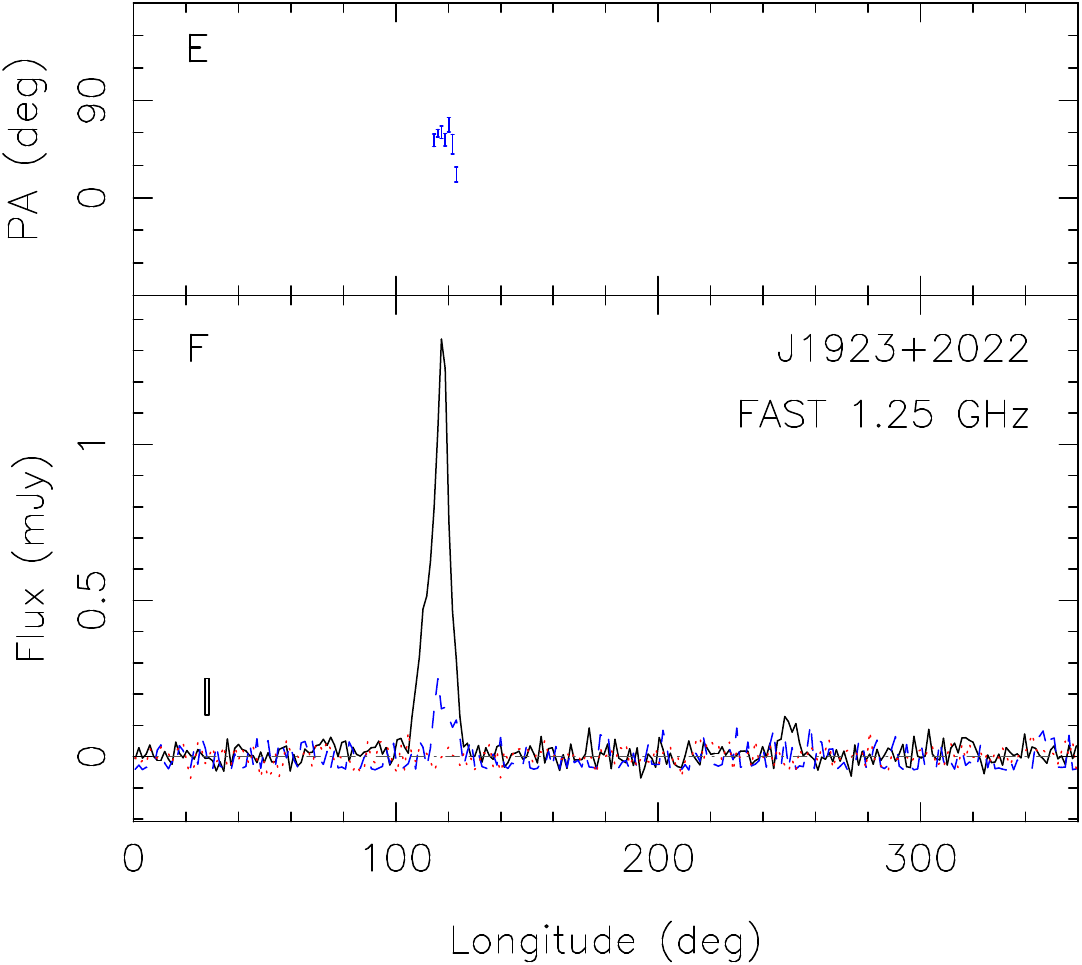}
    \caption{The polarization profiles of PSRs J1851--0108, J1910+0423, and J1923+2022 obtained by adding data from all available FAST tracking observations with a beam offset less than 1.0$'$ offset from the pulsar position, with integration times of 2.3, 1.4, and 0.4 hours, respectively. The total intensity profile (black line), linear (blue dashed line), and circular (red dotted line, positive for the left-hand sense) polarization profiles are plotted in the lower sub-panel, with a scale-box indicating the 1-bin width and the $\pm2\sigma$ of flux density. The linear polarization angles (PA) are plotted in the upper sub-panel with the error bar represented for $\pm1\sigma$. }
    \label{pol}
\end{figure*}

\section{Timing results and discussion}

Using this modified timing model, we follow the methodology of \citet{Freire+2018MNRAS.476.4794F}, and successfully obtained the phase-coherent timing solutions for all three pulsars. We then iterated the data processing procedures to generate improved TOAs and timing, and obtained the final timing results of PSRs J1851--0108, J1910+0423, and J1923+2022 as listed in Table~\ref{basicparameter}, and timing residuals are shown in Fig.~\ref{timing_res}. We searched and did not find optical and infrared counterparts of these systems in  VizieR online data catalogs \citep{vizier} and available survey images including Pan-STARRS1 and UKIDSS \citep{Chambers+2016arXiv161205560C,Lawrence+2007MNRAS.379.1599L}, probably caused by their old ages, large distances and extinction. We analyze the timing results of the three pulsars independently.

\subsection{PSR J1851--0108}

PSR J1851--0108 has a broadened pulse as discussed by \citet{Jing+2025arXiv250614519J} and shown in Fig.~\ref{pol}. It is moving around its companion with an orbital period of 228 days and a mild orbital eccentricity of 0.133. The mass function $f(m_{\rm p},m_{\rm c},\sin i)$ of PSR J1851--0108 gives constraint on the companion mass as
\begin{equation}
\begin{split}
f(m_{\rm p},m_{\rm c},\sin i)&=\frac{m^3_{\rm c}\sin^3 i}{(m_{\rm p}+m_{\rm c})^2}=\frac{4\pi^2}{G}\frac{x^3}{P_{\rm orb}^2}\\
&=0.18562277(14)~M_\odot,
\end{split}
\end{equation}
where $m_{\rm p}$ is the pulsar mass. Assuming a typical pulsar mass of 1.4~$M_\odot$, the companion mass must be greater than 1.03~$M_\odot$. 

The nature of its massive companion can be as follows:   

\begin{enumerate}[(1)]
    \item A massive white dwarf companion. In this case, probably the system has experienced stable Roche-lobe overflow instead of common envelope evolution during the mass transfer from the companion to the neutron star \citep{Ge+2023ApJ...945....7G}, and its mildly eccentric orbit may be induced by the shell flashes when producing its massive white dwarf companion \citep{Tauris+2023pbse.book.....T}. 
    \item A main-sequence star companion. PSR J1851--0108 is a mildly recycled pulsar with a spin period of 87.1 ms and a surface magnetic field strength of $5.0\times10^9$ G, suggesting that this binary pulsar has accreted mass from its companion. Conventional binary stellar evolution models predict neither significant orbital eccentricities nor main-sequence companions in such a wide orbit around a recycled pulsar. However, \citet{Champion+2008Sci...320.1309C} has found an exception, PSR J1903+0327, to this prediction, which may be formed in a hierarchical triple system or a globular cluster. 
    \item A neutron star or black hole companion. In this case, the orbital eccentricity of PSR J1851--0108 should originate from the birth kick of its companion star. However, many studies found that a wide double neutron system with a small orbital eccentricity is less likely to be formed \citep{Tauris+2017ApJ...846..170T,Guo+2024MNRAS.530.4461G,Chen+2025arXiv250814397C}. The birth kicks of black holes are still unclear \citep{Repetto+2012MNRAS.425.2799R,Mandel+2016MNRAS.456..578M,Tauris+2023pbse.book.....T}, but assuming a pulsar mass of 1.4~$M_\odot$, the possibility that the companion mass PSR J1851--0108 is greater than 3.0~$M_\odot$ is only 14\%.
\end{enumerate}

\subsection{PSR J1910+0423}

PSR J1910+0423 has only one main pulse (see Fig.~\ref{pol}). Our timing result shows that PSR J1910+0423 has a spin period of 93.2 ms and a surface magnetic field strength of $9.5\times10^9$ G, indicating that it has been mildly recycled. PSR J1910+0423 and a $\geq0.32~M_\odot$ companion are moving around each other in a 886-day orbit with an orbital eccentricity, therefore this binary system most likely evolved from a low-mass X-ray binary, and the companion star is either a helium white dwarf or a carbon-oxygen white dwarf \citep{Wang+2025RAA....25a4003W}.

\citet{Phinney+1992RSPTA.341...39P} predicted that such a binary pulsar should follow a relationship between its orbital period and orbital eccentricity,
\begin{equation}
    <e^2>^{1/2}\approx1.5\times10^{-4}(P_{\rm orb}/100{~\rm days}).
\end{equation}
The orbital eccentricity of PSR J1910+0423 is 0.0026, close to the expected value of 0.0013.

\subsection{PSR J1923+2022}

PSR J1923+2022 has a main pulse and a discrete secondary pulse with a large rotation longitude separation of about 140$^\circ$. It is also a mildly recycled pulsar that has a spin period of 38.0 ms and a surface magnetic field strength of $2.7\times10^9$ G. Its orbital period is 777 days, and the minimum companion mass is 0.45~$M_\odot$. Similar to PSR J1910+0423, this binary pulsar is also a descendant of low-mass X-ray binaries \citep{Wang+2025RAA....25a4003W}. Its orbital eccentricity of 6.8$\times10^{-4}$ is consistent with the expectation of \citet{Phinney+1992RSPTA.341...39P} as well. The mass function of PSR J1923+2022 is larger than that of PSR J1910+0423, so this binary system should have a larger orbital inclination. \citet{Tauris+1999A&A...350..928T} provides a relationship between the orbital period and the mass of the  white dwarf companion,
\begin{equation}
    \frac{m_{\rm c}}{M_\odot}=(\frac{P_{\rm orb}}{1.1\times10^5{\rm~days}})^{1/4.75}+0.115.
\end{equation}
So the expected companion mass of PSR J1923+2022 is 0.47~$M_\odot$. Compared to the minimum companion mass of 0.45~$M_\odot$,  its orbit is likely close to edge-on.

\begin{figure}[htp]
    \centering
    \includegraphics[width=1.0\linewidth]{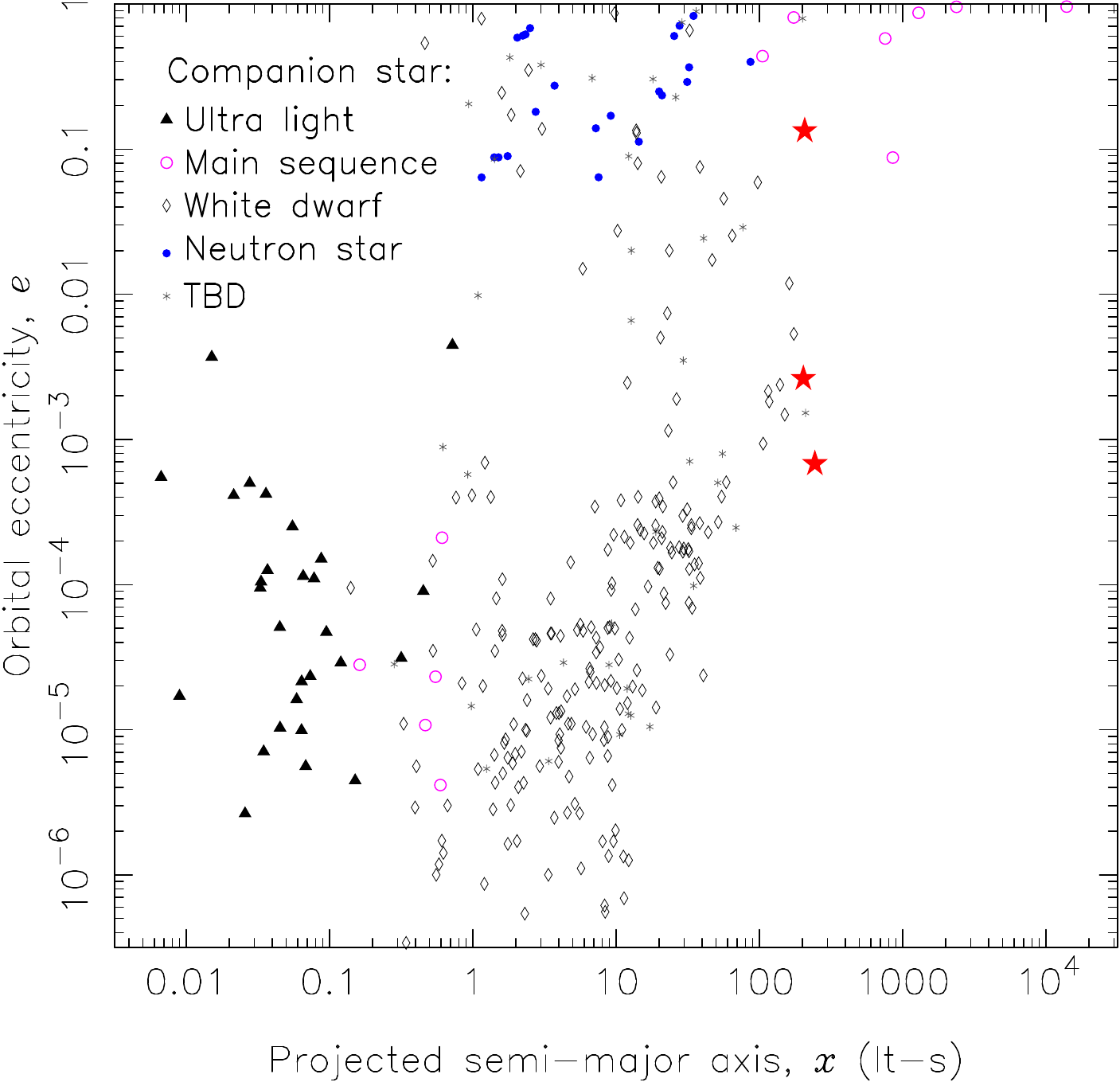}
    \caption{The distribution of projected  semi-major axis $x$ and orbital eccentricity $e$ for binary pulsars. Data from the ATNF pulsar catalog v2.7.0 \citep{Manchester+2005AJ....129.1993M} are shown, with the three GPPS binary pulsars presented in this work highlighted as red stars. }
    \label{x-ecc}
\end{figure}

\section{Summary and discussion}

We report the FAST observations and timing analysis of three binary pulsars—PSRs J1851--0108, J1910+0423, and J1923+2022—discovered in the FAST GPPS survey \citep{Han+2021RAA....21..107H, Han+2025RAA....25a4001H}. These systems exhibit small orbital eccentricities ($e\ll1$) and long orbital periods of 228, 886, and 777 days, respectively (see Fig.~\ref{x-ecc}). 
We developed a modified timing model, designated as \texttt{ELL1R}, based on the \texttt{ELL1} framework \citep{Lange+2001MNRAS.326..274L}. The \texttt{ELL1R} model combines the strengths of both the \texttt{DD} \citep{Damour+1985AIHPA..43..107D, Damour+1986AIHPA..44..263D} and \texttt{ELL1} \citep{Lange+2001MNRAS.326..274L} timing models. 
Using \texttt{ELL1R}, we obtained phase-connected timing solutions for all three pulsars. The timing results indicate that these objects are mildly recycled and likely accreted mass from their companions in the past. While the nature of the companion to J1851–0108 remains uncertain, binary evolution models suggest that PSRs J1910+0423 and J1923+2022 have white dwarf companions and descended from low-mass X-ray binaries. In addition, PSR J1923+2022 is likely in a nearly edge-on orbit.

Its application to the three pulsars demonstrates that: (1) compared to the \texttt{DD} model, \texttt{ELL1R} yields consistent orbital parameters and timing residuals while significantly reducing parameter covariances; (2) compared to the \texttt{ELL1+} model, \texttt{ELL1R} is applicable to a broader range of binary systems and does not require the constraint $xe^4 \lesssim \sigma_{\rm TOA}/N^{1/2}_{\rm TOA}$.

To validate our modified \texttt{ELL1R} timing model, we compared its performance against the \texttt{DD} model \citep{Damour+1985AIHPA..43..107D, Damour+1986AIHPA..44..263D} in \textsc{Tempo2} \citep{Hobbs+2006MNRAS.369..655H} and the \texttt{ELL1+} model \citep{Zhu+2019MNRAS.482.3249Z, Fiore+2023ApJ...956...40F} in \textsc{Pint} \citep{Luo+2021ApJ...911...45L}.

For PSRs J1910+0423 and J1923+2022, the condition $xe^4 \lesssim \sigma_{\rm TOA}/N^{1/2}_{\rm TOA}$ is satisfied, indicating that the low-eccentricity approximation used in \texttt{ELL1+} remains valid. In these cases, the \texttt{ELL1R} model produces timing solutions and residuals consistent with those from \texttt{ELL1+}. In contrast, for PSR J1851--0108, where $xe^4$ reaches 0.065 lt-s, the low-eccentricity approximation breaks down, rendering the \texttt{ELL1+} model inapplicable. Notably, for all three pulsars, the timing residuals obtained with the \texttt{ELL1R} and \texttt{DD} models are identical.

However, the \texttt{DD} model exhibits strong covariances between $T_0$ and $\omega$, leading to significantly larger uncertainties in $T_0$ compared to $T_{\rm asc}$ (see Table~\ref{basicparameter}). This complicates the accurate reproduction of timing results using the reported \texttt{DD} parameters unless more significant digits are retained.

As shown in Eq.~\ref{T0ECCOM}, the \texttt{DD} parameters $T_0$, $e$, and $\omega$ can be derived from $T_{\rm asc}$, $\epsilon_1$, and $\epsilon_2$. Under the assumption of independent errors among $T_{\rm asc}$, $\epsilon_1$, and $\epsilon_2$, the uncertainties of $T_0$, $e$, and $\omega$ can be estimated via the law of propagation of uncertainties. Using this method, we confirmed that the orbital parameters derived from the \texttt{ELL1R} model are consistent with those obtained from the \texttt{DD} model.




\begin{acknowledgments}
This work made use of the data from FAST (https://cstr.cn/31116.02.FAST).  FAST is a Chinese national mega-science facility, operated by National Astronomical Observatories, Chinese Academy of Sciences. 
The authors were supported by the National Natural Science Foundation of China (NSFC), grant numbers 12588202 and 12041303, National Key R\&D Program of China No. 2025YFA161400, the Chinese Academy of Sciences via project JZHKYPT-2021-06, 
%
%
and the National SKA Program of China grant 2020SKA0120100 and 2022SKA0120103.
\end{acknowledgments}





%
\facilities{FAST}

\software{Pint \citep{Luo+2021ApJ...911...45L}; Tempo2 \citep{Hobbs+2006MNRAS.369..655H}; Psrchive \citep{Hotan+2004PASA...21..302H} \textsc{Dspsr} \citep{Straten+2011PASA...28....1V}; Presto \citep{Ransom+2001PhDT.......123R}.
          }




\bibliographystyle{aasjournalv7}



\end{document}